\documentclass[12pt]{article}
\usepackage[mathscr]{eucal}
\usepackage{amssymb,latexsym}
\usepackage{verbatim}
\usepackage{graphicx}
\usepackage{amsmath}
\usepackage{amsthm}
\usepackage{enumerate}
\usepackage{authblk}
\usepackage{color}

\usepackage[margin=.75in]{caption}

\newtheorem{thm}{Theorem}[section]
\newtheorem{lem}[thm]{Lemma}
\newtheorem{Def}[thm]{Definition}
\newtheorem{prop}[thm]{Proposition}
\newtheorem{cor}[thm]{Corollary}

\newtheorem{innercustomthm}{Theorem}
\newenvironment{customthm}[1]
  {\renewcommand\theinnercustomthm{#1}\innercustomthm}
  {\endinnercustomthm}

\renewcommand\l{\lambda}

\newcommand\wt{\widetilde}

\newcommand\s{\sigma}
\renewcommand\d{\partial}

\renewcommand\l{\lambda}
\newcommand\g{\gamma}

\renewcommand\a{\alpha}

\newcommand\e{\epsilon}

\newcommand\beq{\begin{equation}}
\newcommand\eeq{\end{equation}}
\newcommand\ben{\begin{enumerate}}
\newcommand\een{\end{enumerate}}
\newcommand\bit{\begin{itemize}}
\newcommand\eit{\end{itemize}}

\newcommand{\ov}{\overline}

\newcommand{\ext}{\text{{\rm ext}}}
\newcommand{\pd}{\partial}

\newcounter{mnotecount}

\title{Timelike completeness as an obstruction to $C^0$-extensions}

\author[1]{Gregory J. Galloway}
\author[1]{Eric Ling}
\author[2]{Jan Sbierski}

\affil[1]{Department of Mathematics 

University of Miami, Coral Gables, FL, USA }

\affil[2]{Department for Pure Mathematics and Mathematical Statistics

University of Cambridge, Cambridge, UK

}

\begin{document}
\date{}
\maketitle
\vspace{.2in}

\begin{abstract} 
The study of low regularity (in-)extendibility of Lorentzian manifolds is motivated by the question whether a given solution to the Einstein equations can be extended (or is maximal) as a weak solution. In this paper we show that a timelike complete and globally hyperbolic Lorentzian manifold is $C^0$-inextendible. For the proof we make use of the result, {recently established by S\"amann \cite{Samann}}, that even for \emph{continuous} Lorentzian manifolds that are globally hyperbolic, there exists a length-maximizing causal curve between any two causally related points.
\end{abstract}


\section{Introduction}

A smooth Lorentzian manifold is said to be $C^k$-inextendible, if it cannot be isometrically embedded as a proper subset into another Lorentzian manifold of the same dimension with a $C^k$-regular metric. It is a classical result that a smooth timelike geodesically complete Lorentzian manifold is $C^2$-inextendible {(see e.g. \cite[Proposition 6.16]{BEE})}. In this paper we prove

\begin{customthm}{1} \label{Thm1}
A smooth (at least $C^2$) time-oriented Lorentzian manifold that is timelike geodesically complete and globally hyperbolic is $C^0$-inextendible.
\end{customthm}

{This fundamental inextendibility result is an almost immediate consequence of the following.

\begin{customthm}{2} \label{Thm2} Let $(M,g)$ be a smooth (at least $C^2$) time-oriented globally hyperbolic Lorentzian manifold.  If $(M,g)$ admits a  $C^0$-extension,  then it contains a timelike geodesic that has an end point on the boundary of $M$.
\end{customthm}}

The study of extensions of Lorentzian manifolds arises naturally in Einstein's theory of general relativity as the classical question, whether a given Lorentzian manifold, a solution to the Einstein equations, can be extended to a bigger solution. Traditionally, this question has been addressed mainly for extensions with a Lorentzian metric of regularity at least $C^{2}$. The statement that a given solution of the Einstein equations is inextendible as a \emph{Lorentzian manifold} of a certain regularity entails in particular that the solution is inextendible (or `maximal') as a \emph{solution} of the considered regularity. 

For a Lorentzian manifold to be a strong solution of the Einstein equations, the Lorentzian metric has to be at least twice differentiable. For weak solutions to the Einstein equations, one requires that the metric is at least continuous and, in some coordinate system, the Christoffel symbols are locally in $L^2$.
Thus, in order to rule out a continuation of a given solution of Einstein's equations as a weak solution it suffices to show that it is inextendible as a Lorentzian manifold with a continuous metric and Christoffel symbols that are locally square integrable. 

The study of low regularity inextendibility is in particular relevant for the \emph{strong cosmic censorship conjecture} in general relativity. {In a formulation according to Christodoulou (see the prologue of \cite{Chris09}) this states} that for generic asymptotically flat (or compact) initial data for the Einstein equations the corresponding maximal globally hyperbolic development is inextendible as a {weak solution to the Einstein equations. As discussed above, this would follow from showing that for generic asymptotically flat (or compact) initial data for the Einstein equations the corresponding maximal globally hyperbolic development is inextendible as a Lorentzian manifold with a continuous metric and Christoffel symbols locally in $L^2$.} One way to paraphrase this conjecture is to say that general relativity is generically a deterministic theory. For more on this conjecture we refer the reader to the introductions of \cite{Sbierski}, \cite{GalLing}, \cite{Chris09}, \cite{Dafermos03} and references therein.
A further motivation for the study of \emph{low regularity} inextendibility is to give a classification of the strength of the breakdown of the metric, {i.e., the necessary loss of regularity of an extension}. {If the breakdown of the metric signals a `singularity', then this would correspond to a classification of the strength of singularities.}
\newline

A systematic study of low-regularity inextendibility was started in \cite{Sbierski}, which develops some general methods for proving the $C^0$-inextendibility of Lorentzian manifolds and applies them to show in particular the $C^0$-inextendibility of the maximal analytic Schwarzschild spacetime as well as that of the Minkowski spacetime. The work \cite{GalLing} extends these methods and investigates the extendibility properties of the open cosmological FLRW spacetimes, finding that so called `Milne-like' FLRW spacetimes are $C^0$-extendible while establishing the $C^0$-inextendibility of `non-Milne-like' FLRW spacetimes in the spherically symmetric class of extension. In all of the inextendibility results it is important to understand the obstruction to extensions stemming from the region of spacetime which, in a vague sense, is `timelike asymptotically complete'. In the Schwarzschild spacetime this would be the exterior region of the black hole, in Minkowski space the whole spacetime, and in the open cosmological spacetimes the future of any Cauchy hypersurface. The methods developed in \cite{Sbierski} to capture the obstruction to $C^0$-extensions of these regions require  future divergence\footnote{A spacetime $(M,g)$ is said to be \emph{future divergent} iff for every future inextendible timelike curve $\gamma : [0, \infty) \to M$ the timelike distance $d\big(\gamma(0), \gamma(s)\big)$ between $\gamma(0)$ and $\gamma(s)$ goes to infinity for $s \to \infty$.} and future one-connectedness\footnote{{A spacetime $(M,g)$ is said to be \emph{future one-connected} iff any two future directed timelike curves with the same endpoints are homotopic with fixed endpoints via timelike curves.}} (at least some asymptotic form of it) of the spacetime. Especially the property of future one-connectedness, if it holds, is often quite difficult to prove. In this paper we circumvent the property of future one-connectedness by proving that the properties of global hyperbolicity and timelike geodesic completeness also ensure the $C^0$-inextendibility of the spacetime, thus enlarging the available toolbox for proving low-regularity inextendibility.  In particular, Theorem \ref{Thm1} applies directly to the spacetimes arising dynamically from perturbed Minkowski initial data, \cite{ChrisKlainStability}, and from perturbed de Sitter initial data, \cite{Fried86a}, \cite{Fried86b}, \cite{And05}. 
\newline

The proof of Theorem \ref{Thm2} {makes use of the following theorem, which was recently proved in the very nice paper of  S\"amann \cite{Samann}.} 

\begin{customthm}{3}[{\cite[Proposition 6.4]{Samann}}]\label{Thm3}
Let $(M,g)$ be a time-oriented globally hyperbolic Lorentzian manifold with a continuous metric. Let $q \in J^+(p)$. Then there is a causal curve $\g$ from $p$ to $q$ which has length greater than or equal to that of all other causal curves from $p$ to $q$.\footnote{{It is shown in \cite{Samann} that the usual assumption of `strong causality' in the definition of global hyperbolicity can be weakened to  that of `non-total imprisonment'.}}
\end{customthm}  

{The analogue of this theorem for Lorentzian manifolds $(M,g)$ with $g$ at least in $C^2$ is a classical result, see for example \cite{BEE}.  S\"amann's proof is more in the spirit of Seifert's original proof \cite{Seifert} in the smooth setting, which is based on the compactness of $C(p,q)$, the space of causal curves from $p$ to $q$, in a suitable topology.    In Section \ref{Prelim}, we include a slightly more direct proof of Theorem \ref{Thm3} along the lines of \cite[Proposition 14.7]{BEE}, based on limit maximizing curves. Both proofs are ultimately closely related, and require the upper semicontinuity of the length functional in appropriate settings.}


The Riemannian analogue of Theorem \ref{Thm3}, which states that for any two points in a complete Riemannian manifold with a continuous metric, there exists a length-minimizing curve connecting these points, is a well-known textbook result, see for example Theorem 2.5.23 in \cite{BuBuIv01}. Removing the completeness assumption, it still holds that for any point on a Riemannian manifold with a continuous metric one can find a small neighborhood, such that any two points in this neighborhood can be connected by a length-minimizing curve.

{With regard to Theorem \ref{Thm1}}, it is instructive to compare the Riemannian case and the Lorentzian case. Given a complete Riemannian manifold $(M,g)$, it also holds that it is $C^0$-inextendible: assuming there exists a $C^0$-extension, one considers a neighborhood of a boundary point and finds a length-minimizer that connects a point in $M$ to this boundary point. The portion of this curve in $M$ has to be an inextendible geodesic, which by the assumption of completeness has to have infinite length. This gives the contradiction. 

In the Lorentzian case one proceeds analogously, one considers a length-maximizer connecting  the boundary point with a causally related point in $M$. The difference now is, that in order to obtain the contradiction, one has to rule out {the subtle possiblility} that the part of this length-maximizer that is contained in $M$, is a \emph{null} geodesic {(note that this length-maximizer might have non-trivial extent in the complement of $M$)}. It is here that we make use of the global hyperbolicity of $M$.

{The  proofs of Theorems \ref{Thm1}  and \ref{Thm2}} are given in Section \ref{SecInex}, where we also conclude with some applications.

\subsection*{Acknowledgements}

Jan Sbierski would like to thank Magdalene College, Cambridge, for their financial support and the University of Miami for hospitality during a visit when this project was started.  The authors {are grateful to Piotr 
Chru\'sciel, Ettore Minguzzi, and the anonymous referees for some helpful comments.   The authors also thank Clemens S\"amann for bringing to their attention the paper \cite{Samann}.}

\section{{Preliminaries}}\label{Prelim}

In this section we introduce some basic causal theoretic notions relevant to the $C^0$ setting.  Also, for the convenience of the reader, we present a rather basic `barebones'  proof of the existence of maximizers in this setting.

By a  $\mathbf{C^k}${\bf \emph{spacetime}}  we mean a smooth, connected, and paracompact $d+1$ dimensional manifold $M$, equipped with a $C^k$ Lorentzian metric $g$, such that $(M,g)$ is time-orientable. Henceforth, throughout this section, we restrict attention to $C^0$ spacetimes.
 For causal theoretic notions used here, we will {by and large} follow the development of Chru\'sciel \cite{Chru} and Chru\'sciel and Grant \cite{ChruGrant}.  Fix a complete Riemannian metric $h$ on $M$.
A curve $\g: I \to M$ is said to be {\bf \emph{locally Lipschitzian}}
if given any compact $K \subset I$, there is a constant $C(K)$ such that for all $s_1, s_2 \in K$, we have 
\[
d_h\big(\g(s_1),\g(s_2)\big) \leq C(K)|s_1 - s_2|
\]
where $d_h$ is the Riemannian distance function on $(M,h)$.  (As shown in \cite{Chru}, this definition is independent of the choice of the complete Riemannian metric.)  By Rademacher's theorem, a locally Lipschitzian curve $\g$ is differentiable almost everywhere and $\g' \in L^\infty_{\text{loc}}$. A locally Lipschitzian curve $\g: I \to M$ is said to be {\bf \emph{future timelike}} or {\bf \emph{future causal}} if its derivative is future timelike or future causal, respectively, almost everywhere.\footnote{{This means, in particular, that the derivative is nonzero almost everywhere.}} For compact intervals $I$, 
the {\bf \emph{length}} of $\g$ is given by, $L(\g) = \int_I \sqrt{-g(\g', \g')}$.   It is a useful fact, shown in \cite{Chru}, that any causal curve $\g: I \to M$ can be reparameterized with respect to $h$-arc length, so that $\g$ is (uniformly) Lipschitzian in this parameterization.

Let $U$ be an open set about $p$.   $J^+(p,U)$ denotes the set of points $q \in U$ which can be reached from 
$p$ via a future causal curve {which is contained in $U$}.  $J^-(p,U)$ is defined time-dually.  We also write $J^{\pm}(p)$ for $J^{\pm}(p,M)$.
The {\bf \emph{Lorentzian distance}} function is the map $d: M \to [0, \infty]$ defined by $d(p,q) = 0$ if $q \notin J^+(p)$ and $d(p,q) = \sup_\g L(\g)$ if $q \in J^+(p)$, where the supremum is taken over all future causal curves joining $p$ to $q$.

A $C^0$ spacetime $(M,g)$ is {\bf \emph{strongly causal}} if given any $p \in M$ and any neighborhood $U$ of $p$, there is a smaller neighborhood $V \subset U$ of $p$ such that no causal curve intersects $V$ in a disconnected set. ($V$ is said to be causally convex.)
$(M,g)$ is {\bf \emph{globally hyperbolic}} if it is strongly causal and the sets $J^+(p) \cap J^-(q)$ are compact for all $p,q \in M$. 

\smallskip
We now present two lemmas, the first of which establishes some simple estimates {(see also 
\cite[Lemma 2.6.5]{Chru})}.

\begin{lem}\label{lem:estimate} Let $(M,g)$ be a $C^0$ spacetime.  For any $p \in M$ and any $\e > 0$ there exists a coordinate neighborhood $(U, x^0, x^1, ...,x^d)$ centered at $p$ such that for any causal curve $\g:[a,b] \to U$,
\beq\label{estimate}
L(\g) <\e \quad{and} \quad L_h(\g) < \e \,,
\eeq 
where $L_h(\g)$ denotes the length of $\gamma$ with respect to the Riemannian metric $h$.
\end{lem} 

\proof   Choose a coordinate neighborhood {$(V,x^i)$ of $p$} so that $x^0$ is a time function, i.e. so that $x^0$ has past timelike gradient {(cf., \cite[Propositon 1.10]{ChruGrant}, \cite[Lemma 2.4]{Sbierski})}.  Put $u = - \nabla x^0/|\nabla x^0|_g$ and $\nu = g(u, \cdot)$. $u$ is a $C^0$ future directed unit timelike vector field on $V$, and $\hat{h} := g + 2\nu \otimes \nu$ is a $C^0$ Riemannian metric on $V$. Put $m = \inf_{p \in \ov{U}} |\nabla x^0(p)|_g$, {where $U$ is a neighborhood of $p$ with compact closure in $V$}. Then if $\g: [a,b] \to U$ is any future directed $g$-causal curve, 
we have,

\begin{align*}
L(\g) &= \int_a^b \sqrt{-g(\g', \g')} = \int_a^b \sqrt{2 [\nu(\g')]^2 - \hat{h}(\g',\g')}
\\
&\leq \sqrt{2} \int_a^b \frac{g(\nabla x^0, \g')}{|\nabla x^0|} \leq \frac{\sqrt{2}}{m}\int_a^b g(\nabla x^0, \g') \\
&= \frac{\sqrt{2}}{m} \int_a^b [x^0 \circ \g]' = \frac{\sqrt{2}}{m} \Big(x^0(\g(b)) - x^0(\g(a))\Big) \,.
\end{align*}
Clearly, by shrinking U we can achieve the first inequality in \eqref{estimate} for all causal curves $\g$ in $U$.  Making $U$ even smaller, if necessary, there exists a constant $C > 0$ such that $h \le C \hat{h}$ on $U$.    Hence, almost everywhere on $[a,b]$,
$$
h(\g',\g') \le C\hat h(\g',\g') = C(g(\g',\g') +2[\nu(\g')]^2)  \le \frac{2C}{|\nabla x^0|^2}[g(\nabla x^0, \g')]^2  \,.
$$
Taking square roots and integrating, we again see that, by shrinking $U$ further, the second inequality in \eqref{estimate} can be satisfied for all causal curves $\g$ in $U$.\qed

\medskip
Taking limits of causal curves is fundamental to causal theory.
Following Minguzzi~\cite{Minguzzi}, we say that the sequence of future causal curves $\g_n: [0, b_n] \to M$ converges uniformly to 
$\g:[0,b] \to M$ provided (i) $b_n \to b$ and (ii) for every $\e >0$, there is $N >0$, such that for $n > N$ and for all $t \in [0,b] \cap [0,b_n]$, $d_h(\g(t), \g_n(t)) < \e$.  A future causal curve $\g:[0,b] \to M$ is a {\bf\emph{limit curve}} of the sequence $\g_n: [0, b_n] \to M$ if there is a subsequence $\g_m$ that converges uniformly to $\g$. 


\smallskip
We now make use of the limit curve lemma in \cite{ChruGrant}, to obtain the following.  

\begin{lem}\label{limit curve}
Let $(M,g)$ be a strongly causal $C^0$ spacetime and let $K \subset M$ be compact. Suppose $\g_n: [0,s_n] \to K$ is a sequence of future causal curves, parameterized with respect to $h$-arc length, such that $\g_n(0) \to p$ and $\g_n(s_n) \to q$. Then there is a limit curve $\g: [0, s_*] \to K$ 
such that $\g(0) = p$ and $\g(s_*) = q$. 
\end{lem}

\proof {We can extend each $\g_n$ to a future causal curve $\wt{\g}_n: [0, \infty) \to M$, parameterized with respect to $h$-arc length, which is inextendible by \cite[Theorem 2.5.5]{Chru}}.
Then, by \cite[Theorem 1.6]{ChruGrant},  there exists a subsequence $\{\wt{\g}_m\}$ that converges uniformly on compact subsets to a future inextendible causal curve $\wt{\g}: [0, \infty) \to M$.   Since $K$ is compact, we can cover $K$ by a finite number of arbitrarily small causally convex neighborhoods.   Using Lemma \ref{lem:estimate}, we then see that the sequence of $h$-lengths  $\{s_m\}$ is bounded above.
Passing to a subsequence if necessary, we may assume that   
$s_m \to s_* < \infty$.  Hence, by the uniform convergence, $\g(s_*) = \lim_{m \to \infty} \g_m(s_m) =q$. Clearly $\wt{\g}(0) = p$, so let $\g:[0 ,s_*] \to M$ be the restriction of $\wt{\g}$ to the interval $[0,s_*]$. By the uniform convergence, it follows that $\g \subset K$.\qed

\medskip

A key step in the proof of the existence of maximizers is recognizing that the length functional is upper semicontinuous. This was first proved by Penrose \cite{Penrose} for strongly causal $C^2$ spacetimes. It was later observed in \cite{Gal, GalEsch}
that the uniform convergence on compact subsets enables one to prove upper semicontinuity in spacetimes without assuming strong causality.  We now present  a proof of upper semicontinuity for $C^0$ spacetimes; see \cite[Theorem 6.3]{Samann} for a closely related proof in a slightly different setting.

\medskip

\begin{prop}\label{Uppersemi for length}
Let $(M,g)$ be a $C^0$ spacetime. 
Suppose a sequence of future causal curves $\g_n: [a,b] \to M$ converges uniformly to the causal curve
$\g: [a,b] \to M$.  Then $L(\g) \geq \limsup\limits_{n \to \infty} L(\g_n)$. 
\end{prop} 

\proof By Proposition 1.2 of \cite{ChruGrant}, there is a family of smooth Lorentzian metrics $\{g_\e: \e > 0\}$ such that $g_\e$ is wider than $g$ {(i.e., $g(X,X) \le 0$, $X \ne 0 \implies g_{\e}(X,X) < 0$)}, $g_\e$ converges uniformly on compact subsets of $M$ to $g$ as $\e \to 0$, and for all $X \in TM$ with $|X|_h = 1$, we have $|g(X,X) 
- g_\e(X,X)| < \e$.  

Hence, the curves $\g$ and $\g_n$ are future causal curves in $(M,g_\e)$.
By Lemma \ref{estimate}, there exists $C>0$ and a partition $a = s_0 < s_1 < \dotsb < s_k = b$ of $[a,b]$ such 
that, for {$i = 0, ..., k-1$, $\g([s_i,s_{i+1}])$
 lies in a neighborhood $V_i$} with the property that every $g$-causal curve in $V_i$ has $h$-length less 
than $C$. 

Now let $\s$ be a $g$-causal curve in $V$ parameterized by $h$-arclength, then a.e.,
\[
\sqrt{|g(\s', \s')|} < \sqrt{|g_\e(\s', \s')| + \e} < \sqrt{|g_\e(\s', \s')|} + \sqrt{\e}
\]
and so 
\[
L_g(\s) < L_{g_\e}(\s) + L_h(\s)\sqrt{\e} < L_{g_\e}(\s) + C \sqrt{\e} \,.
\]
Switching the roles of $g$ and $g_\e$, we establish that 
\begin{equation}\label{length less than smooth length}
|L_g(\s) - L_{g_\e}(\s)| < C\sqrt{\e}  \,.
\end{equation}
It follows that,
\beq\label{lengthcomp}
L_g(\g) > L_{g_\e}(\g) - Ck\sqrt{\e} \, ,\quad \text{and for large $n$,} \quad L_{g_\e}(\g_n) > L_{g}(\g_n) - Ck\sqrt{\e} \,,
\eeq
{since for large $n$, $\g_n([s_i,s_{i+1}]) \subset V_i$.}

Now we use (\ref{lengthcomp}) along with the fact that the length functional is upper semicontinuous in the smooth spacetime $(M,g_\e)$.  Indeed, Corollary 2.4.11 in \cite{Chru} implies that  Lipschitz causal curves are continuous causal curves as defined in \cite{Minguzzi}.  Upper semicontinuity for Lipschitz curves in smooth spacetimes then follows from \cite[Theorem 2.4(a)]{Minguzzi} (see also \cite{Penrose}). 

Hence, we have,
\begin{align*}
L_g(\g) &> L_{g_\e}(\g) - Ck\sqrt{\e}
\\
&\geq \limsup\limits_{n \to \infty} L_{g_\e}(\g_n) - Ck\sqrt{\e}
\\
&\geq \limsup\limits_{n \to \infty} \big(L_g(\g_n) - Ck\sqrt{\e}\big) - Ck\sqrt{\e}
\\
&= \limsup\limits_{n \to \infty} L_g(\g_n) - 2Ck\sqrt{\e}  \,.
\end{align*}
Since $\e > 0$ was arbitrary, the result follows.\qed

%

\bigskip
\noindent
{\it Remark:}  Proposition \ref{Uppersemi for length} remains valid under slightly more general circumstances.  For example, one may assume that each 
$\g_n$ is defined on an $h$-arc length interval $[0,b_n]$, such that $b_n \to b$; 
see \cite[Theorem 2.4(b)]{Minguzzi}.


\medskip

\begin{thm}\label{existence of maximizers}
Let $(M,g)$ be a globally hyperbolic $C^0$ spacetime. Let $q \in J^+(p)$. Then there is a causal curve $\g$ from $p$ to $q$ which has length greater than or equal to that of any other causal curve from $p$ to $q$ (equivalently, $L(\g) = d(p,q)$).
\end{thm}

\proof Let us first observe that the Lorentzian distance function $d$ is finite-valued. Let $q \in J^+(p)$. To prove that $d(p,q)$ is finite, cover the compact set $J^+(p) \cap J^-(q)$ with a finite number of  
causally convex neighborhoods
$\{V_1, \dotsc, V_k\}$ such that $d_{V_i}$ is bounded by $1$ for all $i = 1, \dotsc, k$ (cf., Lemma~\ref{lem:estimate}).  Any future causal curve $\l$ from $p$ to $q$ can only enter each $V_i$ once so $L(\l) \leq k$. Since $\g$ was arbitrary, we have $d(p,q) \leq k$. 

Now by definition of $d$, there is a sequence of future causal curves $\g_n: [0, s_n] \to M$ from $p$ to $q$, parameterized with respect to $h$-arc length, which satisfy $L(\g_n) \geq d(p,q) - n^{-1}$.  By passing to a subsequence if necessary, we may assume 
by Lemma~\ref{limit curve} that the sequence $\g_n$ converges uniformly to a future causal curve $\g: [0, s_*] \to M$ from $p$ to $q$. Then Proposition \ref{Uppersemi for length} (and the remark following its proof) gives 
\[
L(\g) \geq \limsup\limits_{n \to \infty} L(\g_n) \geq \limsup\limits_{n \to \infty} \big(d(p,q) - n^{-1}\big) = d(p,q) \,.
\]
Thus $\g$ is a maximizer. \qed

\smallskip

We note that the proof of Theorem \ref{existence of maximizers} does not make use of the lower semicontinuity of the Lorentzian distance function; compare, for example, the proof of Proposition 14.7 in \cite{BEE} in the smooth case.  In fact, lower semicontinuity does not hold in the $C^0$ setting.  This may be seen as a result of the `bubbling' phenomena discussed in  \cite{ChruGrant}; see especially 
\cite[Example 1.11]{ChruGrant}.

\section{$C^0$-Inextendibility} \label{SecInex}

In this section we will use the existence of maximizers (Theorem \ref{existence of maximizers}) to prove that timelike geodesically complete globally hyperbolic spacetimes are $C^0$-inextendible.

A $C^0$ spacetime $(M_\ext, g_\ext)$ is a {\bf \emph{$C^0$-extension}} of a spacetime $(M,g)$ if they are of the same dimension and $(M,g)$ embeds {smoothly and} isometrically as a proper subset of $(M_\ext,g_\ext)$. If no $C^0$-extension of $(M,g)$ exists, then we say that $(M,g)$ is {\bf\emph{$C^0$-inextendible}}. 

\begin{Def}
{\rm
Given a $C^0$-extension $(M_\ext,g_\ext)$ of $(M,g)$, we make the following definitions (see \cite{GalLing}).
\begin{itemize}
\item The {\bf \emph{future boundary of $M$}}, denoted by $\pd^+ M$,
is the set of points $p \in \pd M$ such that there exists a smooth
future directed timelike curve $\gamma: [0,1] \to M_\ext$ with $\gamma(1) = p$, $\gamma\big([0,1)\big) \subset M$.

\item The {\bf\emph{past boundary of $M$}}, denoted by $\pd^- M$, is the set of points $p \in \pd M$ such that there exists a smooth future directed timelike curve $\gamma: [0,1] \to M_\ext$ with $\gamma(0) = p$, $\gamma\big((0,1]\big) \subset M$.
 \end{itemize}
}
\end{Def}

\smallskip
We will need to make use of the following basic result, which is proved in \cite{Sbierski} (see also \cite{GalLing}).

\medskip

\begin{lem}\label{boundary nonempty}
Let $\iota: (M,g) \to (M_\ext, g_\ext)$ be a $C^0$-extension. Then $\pd^+ M \cup \pd^- M \neq \emptyset$. 
\end{lem}

\smallskip
{
Therefore, if one can find a $C^0$-extension $(M_\ext, g_\ext)$ of $(M,g)$, then one can also find a smooth timelike curve which leaves $M$ and enters $M_\ext$. In fact, if $(M_\ext, g_\ext)$ is a $C^2$-extension of $(M,g)$, then, by using normal neighborhoods, one can find timelike geodesics which leave $M$ and enter $M_\ext$. The next {theorem} is our key result. It says that when $(M_\ext, g_\ext)$ is a $C^0$-extension of a globally hyperbolic spacetime $(M,g)$, then one can \emph{still} find timelike geodesics which leave $M$ and enter $M_\ext$.


\begin{thm}\label{timelike geodesics leave M}
Let $(M,g)$ be a smooth (at least $C^2$) globally hyperbolic spacetime. Suppose $(M_\ext, g_\ext)$ is a $C^0$-extension of $(M,g)$. If $\pd^+M \neq \emptyset$, then there is a future {directed} timelike geodesic in $M$ {that has a future endpoint on $\partial M \subseteq M_\ext$}.  
\end{thm}

}

\proof Let $p \in \pd^+M$. Let $\g: [0,1] \to M_\ext$ be a smooth future directed timelike curve such that $\g(1) = p$ and $\g\big([0,1)\big) \subset M$. Let $\e > 0$ to be fixed later. By continuity of $g_\ext$ and after a possible reparameterization of $\gamma$, there is a $\delta > 0$ (depending on $\e$) and a coordinate system
 \[
 \phi = (x^0, x^1, \dotsc, x^d): U_\e \to (-4\delta, 4\delta)^{d+1}
 \]
around $p$, where $U_{\e}$ is a relatively compact open subset of $ M_\ext$, such that 
{(cf. \cite[Lemma 2.4]{Sbierski})},
\begin{itemize}

\item[(a)] $x^\mu(p) = 0$.

\item[(b)] $x^0\big(\g(s)\big) = \delta(s - 1)$ and $x^i\big(\g(s)\big) = 0$ for all $i = 1, \dotsc, d$.

\item[(c)] $g_{\mu\nu}(p) = \eta_{\mu\nu}$

\item[(d)] $|g_{\mu\nu}(x) - \eta_{\mu\nu}| < \e$ for all $x \in U_\e$ 

\item[(e)] The negative $x^0$-axis lies inside $M$.

\end{itemize}
Here $\eta_{\mu\nu}$ are the usual components of the Minkowski metric and $g_{\mu\nu}$ are the components of $g_\ext$ with respect to $(x^0, x^1, \dotsc, x^d)$. Note that we still assume the entire negative $x^0$-axis lies inside $M$, but we will be mainly interested in the curve $\g$ (which makes up 1/4 of the negative $x^0$-axis).

\medskip

 Consider the following smooth Lorentzian metrics on $U_\e$
\begin{align*}
\eta^{(2)} &= -\frac{1}{4}(dx^0)^2 + \sum_{i = 1}^d(dx^i)^2
\\
\eta^{(1/2)} &= -4(dx^0)^2 + \sum_{i = 1}^d(dx^i)^2   \,.
\end{align*}

By continuity of the metric, there is an $\e_0 > 0$ such that for any $0 < \e < \e_0$ and any $X \in TU_{\e}$, we have {(cf., \cite[Proposition 1.10]{ChruGrant}},
\begin{align}
\eta^{(2)}(X,X) \le 0  \:\:\:\: &\Longrightarrow \:\:\:\:  g_\ext(X,X) < 0  \label{slope bound 1}
\\
g_\ext(X,X) \leq 0 \:\:\:\: &\Longrightarrow \:\:\:\: \eta^{(1/2)}(X,X) < 0 \label{slope bound 2}  \,.
\end{align}

For $0 < \e < \e_0$ we consider $V_\e \subset U_\e$ which is given by 
\[
V_\e = I^+_{\eta^{(1/2)}}\left(\phi^{-1}\left(-2\delta, 0, \dotsc, 0\right), U_\e\right) \cap I^-_{\eta^{(1/2)}}\big(\phi^{-1}(2\delta, 0, \dotsc, 0), U_\e\big).
\]
We show that $(V_\e,g_\ext|_{V_\e})$ is globally hyperbolic:
First note that, by \cite[Corollary 2.4.11]{Chru},  Lipschitz causal curves in $(V_\e,\eta^{(1/2)})$ can be closely approximated by piecewise smooth causal curves.  Hence, since 
$(V_\e,\eta^{(1/2)})$ is strongly causal and $\eta^{(1/2)}$ is wider than $g_\ext$, it is almost immediate that $(V_\e,g_\ext|_{V_\e})$ is strongly causal. Now fix ${r,s} \in V_\e$ with $s \in J^+_{g_\ext}(r,V_\e)$.  We have to show $D:= J^+_{g_\ext}(r,V_\e) \cap J^-_{g_\ext}(s,V_\e)$ is compact. Since $\eta^{(1/2)}$ is wider than $g_\ext$, it is easy to see that $D$ is a subset of  $J^+_{\eta^{(1/2)}}(r,V_\e) \cap J^-_{\eta^{(1/2)}}(s,V_\e)$, which is clearly compact in $U_\e$.  Hence, it is sufficient to show that $D$ is closed in $V_\e$.  This follows from an application of Lemma \ref{limit curve}.

By Theorem \ref{existence of maximizers}, there is a maximizer $\alpha: [0,2] \to V_{\e}$ from $q := \g(0) = \phi^{-1}(-\delta, 0, \dotsc, 0)$ to $p = \g(1)$. Since $\a$ is a maximizer, we know that 
\[
L(\a) \geq L(\g).
\] 
Now $\alpha$ begins in the physical spacetime $(M,g)$ and ends at $p \in \pd^+M$.  Therefore, we can break $\alpha$ into two curves $\s: [0,1) \to V_\e$ and $\l: [1,2] \to V_\e$ where 
\begin{itemize}

\item[(1)] $\s(s) = \alpha(s) \:\:\: \text{ for } \:\:\: s \in [0,1)$

\item[(2)] $\l(s) = \alpha(s)\:\:\: \text{ for } \:\:\: s \in [1,2]$

\item[(3)] $\s\big([0,1)\big) \subset M$

\item[(4)] $\l(1) \in \pd M$.

\end{itemize} 

Since $\alpha$ is a maximizer in $V_\e$, we know that $\s$ is a maximizer in $V_\e$. Since $\s$ lies inside $M$,  {it follows from \cite[Proposition 34, p.\ 147]{ON} (which can be extended to Lipschitz causal curves 
using \cite[Proposition 2.4.5]{Chru})}  that, up to parametrization, $\s$ is either a timelike or a null geodesic. 
{(In the special case $\alpha\big([0,2)\big) \subset M$, it's easy to conclude that $\alpha|_{[0,2)}$ is a timelike geodesic with endpoint $p$.)}

 {Thus it suffices to show that $\s$ cannot be a null geodesic. So let us suppose that $\s$ is a null geodesic. We are going to obtain a contradiction by showing that, for sufficiently small $\e$, we will have 
\[
L(\lambda ) = L(\alpha) < L(\g).
\]} 
 


We will do this by putting a lower bound on $L(\g)$ and an upper bound on $L(\lambda)$. To simplify the estimates, we will use the fact that $\g$ and $\l$ can be reparameterized with respect to the $x^0$ coordinate. We can do this since for a small enough neighborhood, $x^0$ is a time function (cf. \cite{Chru}).

\medskip


\noindent\underline{Lower Bound on $L(\g)$}:
\medskip

\noindent With respect to the $x^0$-parameterization, we have $\g(t) = \phi^{-1}(t,0,\dotsc, 0)$. Therefore 
\begin{align*}
L(\g) &= \int_{-\delta}^0 \sqrt{-g_\ext\big(\g'(t),\g'(t)\big)}dt
\\
&= \int_{-\delta}^0 \sqrt{-g_{00}\big(\g(t)\big)}dt
\\
&\geq \delta \sqrt{1 - \e}.
\end{align*}

\noindent\underline{Upper Bound on $L(\lambda)$}:

\noindent Let us write $\lambda(t) = \phi^{-1}\big(t, x^1(t), \dotsc, x^d(t) \big)$. Let $-a \in (-\delta, 0)$ denote the starting parameter value of $\lambda$  with respect to the $x^0$-parameterization. Then 
\begin{align*}
L(\l) &= \int_{-a}^0\sqrt{-g_\ext\big(\l'(t),\l'(t)\big)}dt
\\
&= \int_{-a}^0 \bigg[- g_{00}\big(\l(t)\big) - 2 \sum_{i = 1}^dg_{0i}\big(\l(t)\big)\dot{x}^i(t)
\\
&\:\:\:\:\:\:\:\:\:\:\:\:\:\:\:\:- 2\sum_{1 \leq i < j \leq d}g_{ij}\big(\l(t)\big)\dot{x}^i(t)\dot{x}^j(t) - \sum_{i = 1}^dg_{ii}\big(\l(t)\big)|\dot{x}^i(t)|^2\bigg]^{1/2}dt
\\
&\leq \int_{-a}^0 \left[(1+\e) +2 \sum_{i =1}^d\e|\dot{x}^i(t)| +2 \sum_{1 \leq i < j \leq d}\e|\dot{x}^i(t)\dot{x}^j(t)| \right]^{1/2}dt
\\
&\leq \int_{-a}^0 \big[(1 + \e) + 4d\e + 4d(d-1)\e \big]^{1/2}dt
\\
&= a\sqrt{1 + \e + 4d^2\e}
\end{align*}

\noindent The third line follows from $|g_{\mu\nu}(x) - \eta_{\mu\nu}| < \e$ for all $x \in V_\e$ and the fourth line follows from the fact that $|\dot{x}^i(t)| < 2$, which is a consequence of the bound (\ref{slope bound 2}).

\medskip

Now the bounds on $L(\l)$ and $L(\g)$ and the fact that $L(\l) \geq L(\g)$ gives 
\[
a\sqrt{1 + \e + 4d^2\e} \geq \delta \sqrt{1 - \e}.
\]
Equivalently,
\[
\frac{a}{\delta} \geq \frac{\sqrt{1 - \e}}{\sqrt{1 + \e + 4d^2\e}}.
\]
Now {the aim}
is show that there is a constant $C < 1$ such that $a/\delta \leq C$ and that this constant $C$ is independent of $\e$. Then by choosing $\e$ small enough, we will contradict the above inequality. 

\medskip

\begin{figure}
\begin{center} 
\includegraphics[width=3.8in]{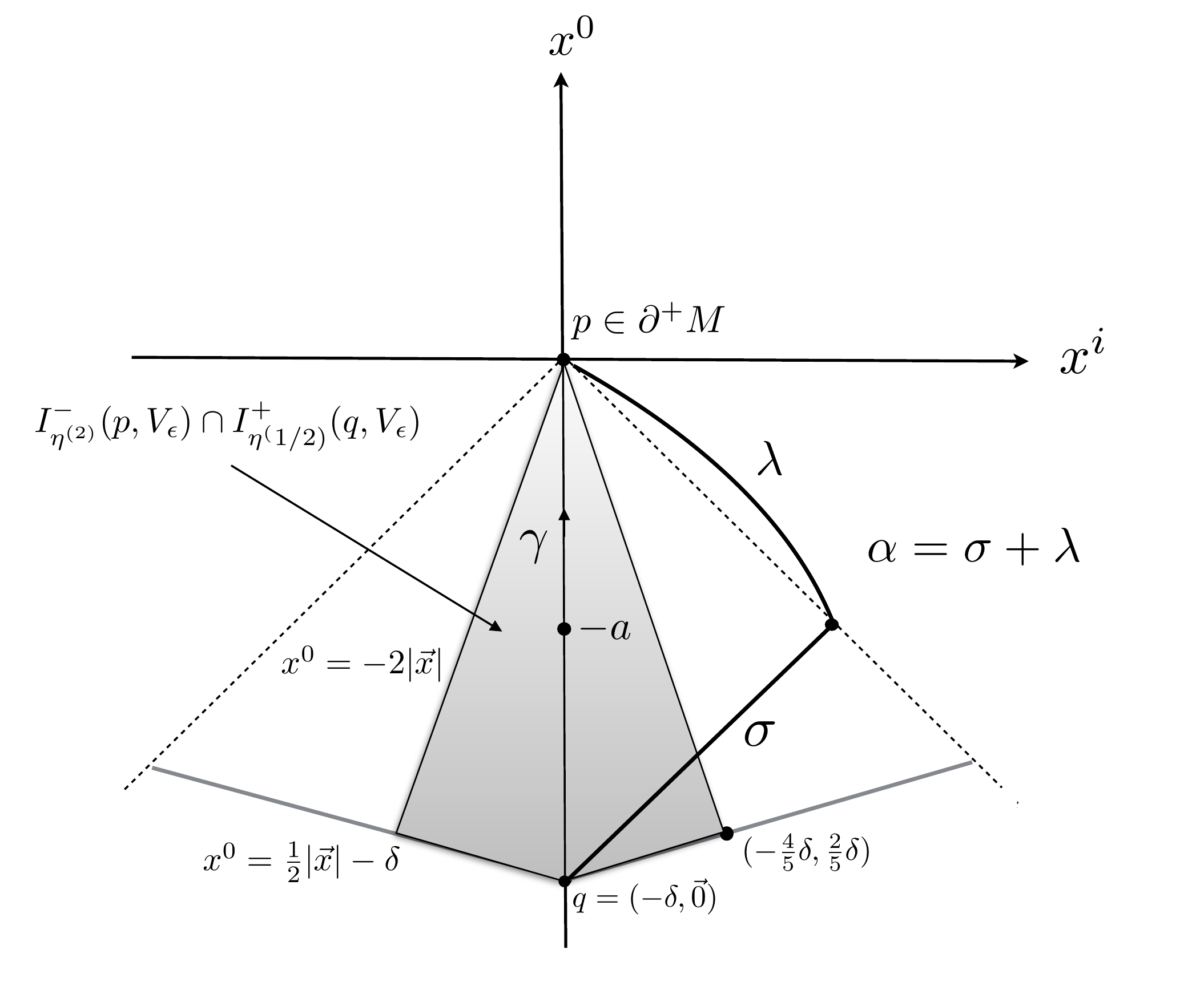}
\end{center} 
\caption[2in]{\small 
\sl $\l(-a)$  lies to  the future of the wide cone and is on or outside the
``narrow" cone.\label{fig:cones} }
\end{figure}

\noindent\underline{Finding $C$}:

\medskip

\noindent With respect to the $x^0$ parameterization, $\l(-a)$ is the starting point of $\l$. By the bounds (\ref{slope bound 2}), we have $\l(-a) \in I^+_{\eta^{(1/2)}}(q,V_\e)$. Recall $q = \phi^{-1}(-\delta, 0, \dotsc, 0)$. We claim that $\l(-a) \notin I^-_{\eta^{(2)}}(p,V_\e)$. Let's assume this to be true for the moment. {We then have $\lambda (-a) \notin I^-_{\eta^{(2)}}(p,V_{\epsilon}) \cap I^+_{\eta^(1/2)}(q,V_{\epsilon})$, and it follows from elementary geometry, see Figure \ref{fig:cones}, that $- a \geq -\frac{4}{5} \delta$.}
That is, $a \leq (4/5)\delta$, and so we can take $C = 4/5$. Note that this $C$ value works for all $0 < \e < \e_0$. So we obtain our contradiction by taking $\e > 0$ small enough so that 
\[
\frac{\sqrt{1 - \e}}{\sqrt{1 + \e + 4d^2\e}} > \frac{4}{5}.
\]

\medskip

\noindent\underline{Last Step. Proving $\l(-a) \notin I^-_{\eta^{(2)}}(p,V_\e)$}:

\medskip

\noindent This is the only place where we will use that $(M,g)$ is globally hyperbolic. {Recall that $\l(-a) \in I^+_{\eta^{(1/2)}}(q,V_\e)$, which in particular implies $x^0\big(\l(-a)\big) > - \delta$. Hence it suffices to show $\l(-a) \notin I^-_{\eta^{(2)}}(p,V_\e) \cap \{\phi^{-1}\big(\{ x_0 > -\delta\}\big) \}$. The proof relies on the following lemma.}

\begin{lem} \label{Lem}
Let $a_0$ and $b_0$ be any points on the negative $x^0$-axis, with $a_0$ to the past of $b_0$. Then 
\[
I^+_{\eta^{(2)}}(a_0,V_\e) \cap I^-_{\eta^{(2)}}(b_0,V_\e) \subset M.
\]
\end{lem}

In particular, the lemma implies 
\begin{equation*}
\begin{split}
I^-_{\eta^{(2)}}&(p,V_\e) \cap \Big(\bigcup_{- \delta > s_1 > -2\delta} I^+_{\eta^{(2)}}(\phi^{-1}(s_1, 0 \ldots, 0),V_\e) \Big) \\
&= \Big(\bigcup_{0 > s_0 > -\delta} I^-_{\eta^{(2)}}(\phi^{-1}(s_0, 0 \ldots, 0),V_\e) \Big)
\cap \Big(\bigcup_{- \delta > s_1 > -2\delta} I^+_{\eta^{(2)}}(\phi^{-1}(s_1, 0 \ldots, 0),V_\e) \Big)\\ &\subseteq M \;.
\end{split} 
\end{equation*}
By elementary geometry we have 
\begin{equation*}
I^-_{\eta^{(2)}}(p,V_\e) \cap \Big{\{}\phi^{-1}\big(\{ x_0 > -\delta\}\big) \Big{\}} 
\subseteq I^-_{\eta^{(2)}}(p,V_\e) \cap \Big(\bigcup_{- \delta > s_1 > -2\delta} I^+_{\eta^{(2)}}(\phi^{-1}(s_1, 0 \ldots, 0),V_\e) \Big)\;,
\end{equation*}
from which, together with $\l(-a) \in \pd M$, the claim follows. This concludes the proof of Theorem \ref{timelike geodesics leave M}. \qed

\medskip

\proof[Proof of Lemma \ref{Lem}:] For technical reasons, it is better to work with the set $M_1$, where $M_1$ is the connected component of $M \cap V_\e$ which contains the negative $x^0$-axis. Each connected component of the intersection of two globally hyperbolic {sub-spacetimes} is globally hyperbolic, so we know that $(M_1, g|_{M_1})$ is globally hyperbolic. It suffices to prove
\[
I^+_{\eta^{(2)}}(a_0,V_\e) \cap I^-_{\eta^{(2)}}(b_0,V_\e) \subset M_1.
\]

Suppose this were not the case. Then there is a point $c_0 \in I^+_{\eta^{(2)}}(a_0,V_\e) \cap I^-_{\eta^{(2)}}(b_0,V_\e)$ while $c_0 \notin M_1$. Let us construct two `straight' lines $\alpha$ and $\beta$ from $a_0$ to $c_0$ and $b_0$ to $c_0$, respectively. Here straight means with respect to the usual Euclidean metric $\delta_{\mu\nu}dx^\mu dx^\nu$. The points $a_0$, $_0b$, and $c_0$ form a two-dimensional triangle with sides given by the negative $x^0$-axis from $a_0$ to $b_0$, $\alpha$, and $\beta$. Let us call this triangle $\Delta$. By rotating coordinates, we can assume that $\Delta$ lies in the $(x^0, x^1)$-plane and that $\Delta$ lies in $x^1 \geq 0$.

Consider the vertical line segments which join $\alpha$ to $\beta$ while keeping $x^1$-constant. Let $T(x^1)$ denote any one of these vertical lines, so that $T(x^1)$ foliates $\Delta$ with parameter $x^1$. Note $T(0)$ is just the negative $x^0$-axis from $a_0$ to $b_0$. Let us define
\[
x^1_* = \sup\{x^1 \mid T(x^1) \subset M_1\}.
\]
Since $T(0) \subset M_1$ is compact, we know that $x^1_* > 0$. Moreover, since $c_0 \notin M_1$, we know that $x^1_* \leq \big(x^0(b_0) - x^0(a_0)\big)/4$. 
Thus there is some point $c_* \in T(x^1_*)$ such that $c_* \notin M_1$. However $T(x^1) \subset M_1$ for all $x^1 < x^1_*$. Therefore we can generate a sequence of points $c_n \in T(x^1_* - 1/n)$ whose only accumulation point is $c_*$. Thus $J^+(a_0,M_1) \cap J^-(b_0,M_1)$ is not compact which contradicts $M_1$ being globally hyperbolic.\qed

\smallskip


{\proof[Proof of Theorems 1 and 2 in the introduction] Lemma \ref{boundary nonempty} and Theorem \ref{timelike geodesics leave M} (along with its time dual) together imply Theorem \ref{Thm2}. Now assume $(M,g)$ is timelike geodesically complete and globally hyperbolic. If $(M_\ext, g_\ext)$ is a $C^0$-extension, then by Theorem 2 there exists a timelike geodesic $\g$ in $M$ with endpoint on the boundary. Since $\g$ has infinite length, this contradicts Lemma \ref{lem:estimate}. {\qed}
}

\smallskip

%
%
%
%

\smallskip
{It is perhaps worth noting that the full strength of global hyperbolicity is not needed in 
Theorem \ref{Thm1}.  Indeed, nowhere in the proof of Theorem \ref{timelike geodesics leave M} is strong causality used, only the compactness of `causal diamonds'. 

Theorem \ref{Thm1} directly applies to the spacetimes arising in the proof of the stability of Minkowski and de Sitter space, by virtue of their global hyperbolicity and timelike completeness:

\smallskip

\begin{cor} \mbox{}  
\begin{enumerate}  
\item[(1)]
The spacetimes constructed in \cite{ChrisKlainStability} which arise dynamically from perturbed Minkowski initial data are $C^0$-inextendible.
\item[(2)]  
The spacetimes constructed in \cite{Fried86a}, \cite{Fried86b}, \cite{And05} which arise dynamically from perturbed de Sitter initial data are $C^0$-inextendible.
\end{enumerate}
\end{cor}

\smallskip

We also note that Theorem \ref{timelike geodesics leave M} implies the following. 

\begin{thm}\label{future is empty}
Let $(M,g)$ be a smooth (at least $C^2$) globally hyperbolic spacetime. Suppose $(M_\ext, g_\ext)$ is a $C^0$-extension of $(M,g)$. If $(M,g)$ is future timelike geodesically complete, then $\pd^+M = \emptyset$. 
\end{thm}

\smallskip

{For globally hyperbolic, future timelike geodesically complete spacetimes, we then have the following structural result for $\d^-M$.} 

\smallskip

\begin{cor}
Let $(M,g)$ be a smooth (at least $C^2$) globally hyperbolic spacetime which is future timelike geodesically complete. Suppose $(M_\ext, g_\ext)$ is a $C^0$-extension of $(M,g)$. Then $\pd^-M$ is an achronal (with respect to smooth timelike curves) topological hypersurface.
\end{cor}

\proof {By Theorem \ref{future is empty}, we know that $\pd^+M = \emptyset$.  The corollary is then an immediate consequence of Theorem 2.6 in \cite{GalLing}.}\qed

\smallskip
{We note further that,  since Theorem \ref{future is empty} avoids the future one-connectedness assumption, {Theorem 3.2 in \cite{GalLing} now extends} to FLRW type models with compact Cauchy surfaces.}

\bibliographystyle{amsplain}
\bibliography{maximizer}

\end{document}